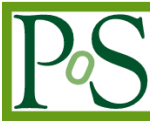

# Proton Beam Intensity Upgrades for the Neutrino Program at Fermilab

## C. M. Bhat[1]

*Fermi National Accelertor Laboratory*
*P. O. Box 500, Batavia IL 60510, USA*
E-mail: `cbhat@fnal.gov`

Fermilab is committed to upgrading its accelerator complex towards the intensity frontier to pursue HEP research in the neutrino sector and beyond. The upgrade has two steps: 1) the Proton Improvement Plan (PIP), which is underway, has its primary goal to start providing 700 kW beam power on NOvA target by the end of 2017 and 2) the foreseen PIP–II will replace the existing LINAC, a 400 MeV injector to the Booster, by an 800 MeV superconducting LINAC by the middle of next decade, with output beam intensity from the Booster increased significantly and the beam power on the NOvA target increased to <1.2 MW. In any case, the Fermilab Booster is going to play a very significant role for the next two decades. In this context, we have recently developed and commissioned an innovative beam injection scheme for the Booster called "early injection scheme". This scheme is already in operation and has a potential to increase the Booster beam intensity from the PIP design goal by a considerable amount with a reduced beam emittance and beam loss. In this paper, we will present results from our experience from the new scheme in operation, current status and future plans.

---

[1]Speaker

1. Introduction

Nearly for twenty-five years since mid-1990 the Fermilab has been the highest energy collider facility in the world providing unprecedented ppbar luminosity for two collider detectors to conduct research in HEP. During this period two of the most important Standard Model particles, top quark and $\nu_\tau$ were discovered along with many precision measurements were made and established hints for Higgs boson. As the ppbar collider program started coming to a conclusion and the LHC at CERN started coming online the Fermilab started focusing research on the neutrino sector by upgrading its facility into a world class accelerator based intensity frontier lab.

Currently, the intensity frontier facility at Fermilab consists of a chain of accelerators: a 750 kV RFQ, 400 MeV normal conducting RF LINAC (200MHz), 0.4-8 GeV (h=84) RCS Booster, 8 GeV (h=588) permanent magnet beam storage synchrotron (Recycler Ring), 8-120 GeV (or 150 GeV) (h=588) Main Injector and 8 GeV Debuncher Ring. The primary goal of the upgrades to these accelerators is to develop a capability of delivering proton beam to i) the NuMI/NOvA target at 120 GeV at the rate of 700 kW, ii) the low energy Booster neutrino experiments at 8 GeV, iii) fixed target precision measurements like g-2 and Mu2e by the end of 2018, and iv) detector development facilities, simultaneously. This stage of the upgrades is ongoing under Proton Improvement Plan (PIP) [1]. The next stage of the upgrade is PIP-II [2] which enables us to increase beam power on the NuMI/NOvA target from 700 kW to 1.2 MW by replacing the existing LINAC by a CW capable superconducting LINAC and supports not only the long-term HEP research programs but forms a platform for the future of the Fermilab. In any case, the Booster is going to play a very important role at least for the next two decades. In this paper we emphasise on the possible improvements in the Booster which would help improving its performance beyond PIP design goals.

Booster is a 15 Hz RCS with a sinusoidal dipole-magnetic ramp, capable of delivering beam continuously on all of its cycles. Until very recently the Booster provided beam at a rate of about 7 Hz. After considerable improvements to the Booster sub-systems we have achieved 15 Hz beam operation with ≈95% acceleration efficiency at about 4.5E12 protons per Booster cycle (ppBc) and longitudinal emittance (LE(95%)) of ≈0.1 eVs/bunch and transverse emittance ≈14πmm-mr. Primarily goal of the Booster upgrade is to increase the beam intensity output further to the downstream facilities at lower beam emittances and at reduced beam loss.

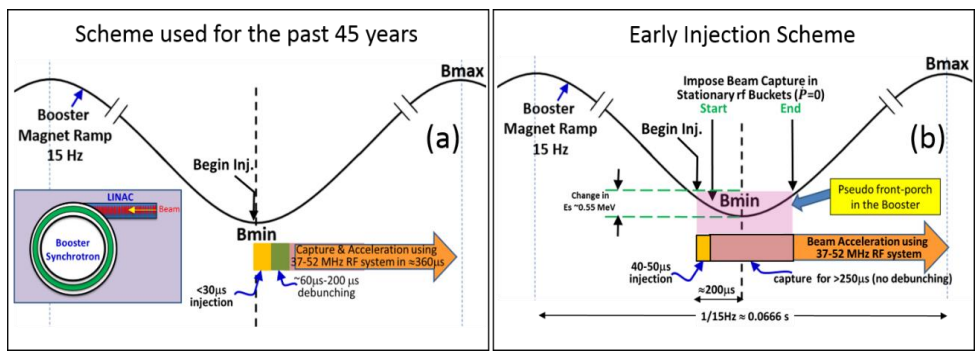

Figure 1: Schematic of the beam injection schemes in to the Booster used a) for the past 45 years, (old scheme), b) recently commissioned early injection scheme. The later scheme can accommodate much longer LINAC beam pulse than the former scheme.

Issues related to the longitudinal beam dynamics has been investigated here. The beam from the LINAC arrives at the Booster with ≈200 MHz bunch structure. Until the end of 2015 the Booster received the beam at the minimum (or close to minimum) of its magnetic field, $B_{min}$, as shown in Fig. 1(a). The injected beam was debunched for 60-200 µsec and subsequently captured quickly (non-adiabatically) using a considerably large RF bucket with particles outside the RF bucket seperatrix leading to some capture loss and LE dilution even before acceleration starts. To accelerate such large bunches it was inevitable of having large RF voltage for the rest of the cycle and > 30 µsec LINAC pulse injection into the Booster was severly limitted by high beam loss at capture. The observed early loss in these cases resembled space charge related issue. To alleviate the capture loss and LE dilution a new scheme is proposed [3]. As shown in Fig.1(b), if we inject beam on the deceleration part of the cycle there will be ample of time to capture beam adiabatically. At the end of 2015 we have implemented the new scheme in operation. Here, we explain briefly the general principle of the scheme, the results from beam dynamics simulations and beam studies, the current status of the scheme in operation and future prospects.

2. New Beam Injection Scheme and Simulations

The basic idea of the new scheme is to inject and start capturing the beam well before *Bdot* = 0 on a *pseudo front porch* created by imposing *dP/dt* = 0 on a changing magnetic field ramp around $B_{min}$. Changing *B* field at a constant momentum still introduces varying revolution frequency in accordance with $\Delta B/B = \gamma_T^2 \, \Delta f / f$ where $\gamma_T$=5.478 is the transition $\gamma$ for the Booster. We demand beam injection at about 200 µsec prior to $B_{min}$ which corresponds to $(\Delta f (RF))_{max}$ ≈15.1 kHz. Thus, on the deceleration part of the ramp the required RF frequency decreases initially and reaches its minimum at $B_{min}$. and increases symmetrically. It is quite important to take the RF frequency variation into account while capturing the beam.

We have demonstrated the feasibility of the scheme in the Booster using 2D- particle tracking simulation code ESME [4] including the longitudinal-space-charge effects for beam intensities in excess of 1E13ppBc. Simulations are performed from injection to beam extraction using the measured beam energy spread at injection, $\Delta E(99\%)$ ≈ 1.25 ±0.20 MeV [5]. A typical case of simulation is depicted in Fig. 2. $H^-$ beam from the LINAC arrives in the form a ribbon (of pulse length in the range of 30-50 µsec) for multi-turn injection into the Booster and proton beam is accumulated after removing the electrons. Figures 2(a) and (b) show the predicted beam particle distributions at about 28 µsec after injection ($T_{rev}$≈2.217 µsec) and at the end of beam capture, respectively, for a case with pre-produced notch (notch width ≈80 ns) upstream of the Booster [6]. Before the completion of the beam capture some part of the pre-produced notch in the Booster is contaminated by protons with high momentum spread. The simulations showed that a) such contamination is less than 0.2% of the total beam in the Booster, b) one can eliminate LE dilution by beam capture in < 240 µsec on the aforementioned pseudo front porch, c) one can keep the LE dilution <10% during acceleration from the injection to the transition crossing, d) keep LE dilution <40% during transition crossing by adding a vernier phase jump of about 10 deg at the transition crossing in addition to a normal phase jump from θ to π-θ and e) snap bunch rotation in the Booster with h=84 RF system for bunches with LE<0.1 eVs (or a combination of 1st and 2nd harmonic RF systems for bunches with LE≤0.17 eVs) can provide the required quality beam to the Recycler for slip stacking.

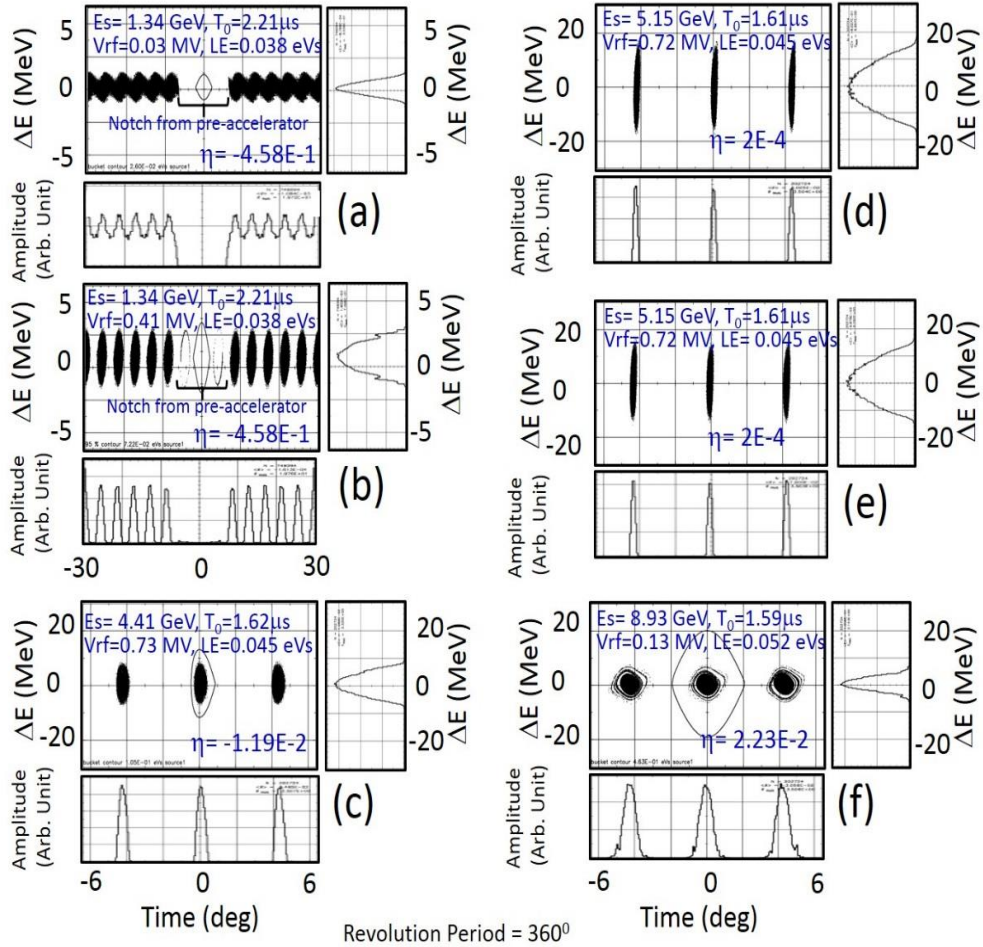

Figure 2: Simulated phase space distributions for a) injection for 1/6 of the Booster ring, b) completion of beam capture, c) about 14 ms into the beam acceleration but below transition energy, d) just after transition crossing, e) after transition crossing with a vernier phase jump of ~6 deg and f) just before beam extraction. The predicted line-charge, energy distribution of the beam particles and LE(95%) per (h=84) bunch are also shown.

We tested the scheme in the Booster following the steps suggested by the simulations and implemented in operation for HEP beam delivery.

### 3.  Beam Studies, Operational Implementation, Issues and Mitigation

Beam tests have been carried out using a specially devised *sequencer* since the newly proposed scheme demanded accelerator operational parameters that are not compatible with the old scheme. Many of the predicted features have been observed during these tests. In addition, some new issues have also been identified and diagnosed which are common to PIP and PIP-II programs, and efforts are in progress to mitigate them. Results from some of the early tests with the new scheme have been reported elsewhere [3].

Late 2015 to mid-2016 many upgrades have been added to the Booster operation (in addition to beam delivery at a rate of 15 Hz) – 1) operational implementation of the above mentioned new scheme, 2) reduce the time jitter between beam event and $B_{min}$ by nearly half, 3) replacing 45-year old time based transition phase-jump by the RF frequency based phase-jump and 4) added a few more 53 MHz RF cavities to handle higher beam intensities in the Booster.

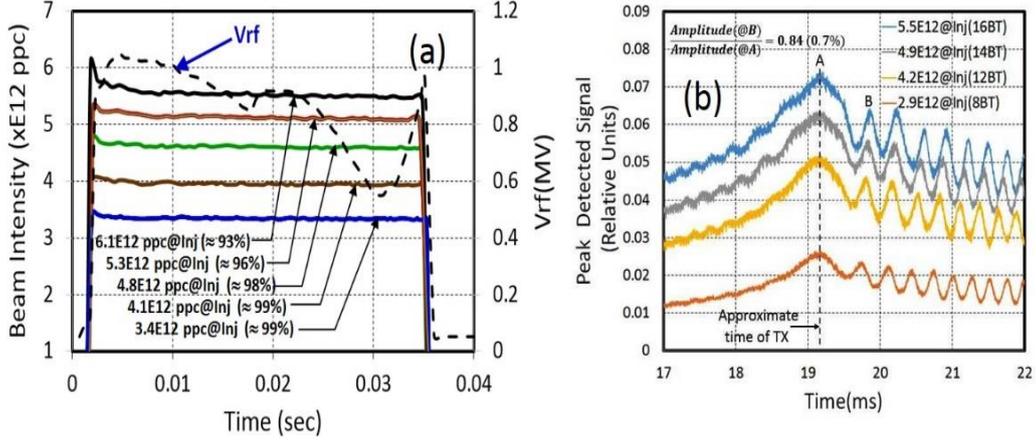

Figure 3: a) Measured beam intensity, transmission efficiency and RF voltage (dashed curve) through the Booster cycle with early injection scheme and frequency based transition crossing. b) Peak detected signal for beam bunches.

Figure 3 displays data from recent beam measurements for intensities in the range of 3.4E12 – 6.1E12 ppBc at injection. The initial step loss in the beam data is due to the notch produced in the Booster. The transmission efficiencies indicated in Fig. 3(a) are corrected for these step losses. The dashed curve in Fig 3(a) is the sum of the measured cavity fan-back RF voltage. We observed that the total RF power per Booster cycle for the new scheme is about 15% smaller than that used while the old scheme was in operation. The frequency based transition crossing made the transition crossing much robust and eliminated the randomness in time of transition phase jump [7]. The amplitude ratios between 1st and 2nd peak of the observed oscillations in the data shown in Fig. 3(b) found to be fairly constant, which is an indicative of reduced importance of longitudinal space-charge effect at the beam intensities up to about 6.3E12 ppBc. The simulations showed that the observed oscillations in the peak detected signals are mainly due to bunch to bucket phase mismatch.

We have measured the longitudinal emittance, LE (95%) at the end of the beam capture and close to the beam extraction. The average LE at the end of capture ~0.05±0.01 eVs while at extraction varies in the range of 0.1 -0.14 eVs (±10%) with bunch intensity. The LE dilution from injection to the end of the capture is ≈ 50%. We do not see any beam losses in the early part of the cycle and the observed LE dilution can be minimized by optimizing the RF voltage curve used during the capture. The emittance growth from the end of capture to the end of the cycle is mainly arising during the transition crossing and can be minimised by adding a vernier phase jump at transition crossing (mentioned earlier in Section 2). Currently, capability of addition of such phase adjustment in operation is in progress.

The early injection scheme primarily addresses issues related to the longitudinal beam dynamics and helps improving transverse dynamics by providing lower $\Delta E$ beam through the cycle. The Booster has 24 symmetric sectors with independently controllable higher ordered correctors (quads, skew quads, sextupoles and skew-sextupoles) to control its transverse beam dynamics from injection to extraction. In the past we used identical ramps on each family of magnets, while for the last several years we have added individually controlled offsets to these ramps. Additional efforts to mitigate the observed beam loss for the first 4-5 msec after the beam capture (which is believed to be due to the transverse beam dynamics) by utilizing the ability of independently controllable magnet correctors are in progress.

Ever since the early injection was put in operation we were able to inject up to about 44 μsec long LINAC beam pulse into the Booster (which is <30% than that in the past) and used it for Neutrino Experiments.

### 4.     Conclusions and Future Prospects

The Fermilab Booster is going to remain as one of the centre-pieces in the on-going upgrade program and necessary improvements need to be made to handle the increased beam power from PIP-II. Over the years we have identified and mitigated a number of problems related to beam dynamics issues. However, the method of beam injection and capture in use for the past 45 years had a serious limitation to provide beam to users on a regular basis with increased LINAC pulse length > 30 μsec with reduced the beam loss at the same time. In this regard, a new scheme is developed which enabled us to inject the beam on the deceleration part of the dipole magnet ramp at about 200 μsec prior to the $B_{min}$ on a pseudo front porch. The start of the beam acceleration is imposed to commence only after the beam is fully captured adiabatically in longitudinal phase space. We demonstrated that the scheme in operation helped to reduce/eliminate beam loss during the early part of the beam cycle and at the same time 1) increase the beam intensities from about 4.6E12 ppBc to > 5.7E12 ppBc, with an acceleration efficiency over 93% and 2) with an acceptable longitudinal emittance even at these higher beam intensities. Thus, this scheme along with mitigation of the issues mentioned above has a potential to increase the Booster beam output by nearly 35% from PIP design goal and will help a smooth transition from PIP to PIP-II.

**Acknowledgement**

Author would like to thank W. Pellico, S. Holmes, C. Drennan, K. Triplett, S. Chaurize, K. Seiya, B. Hendricks, T. Sullivan, F. Garcia, C. Tan and A.Waller for many useful discussions and help in the beam studies. Special thanks are due to Fermilab Accelerator Division MCR crew.